\documentclass{article}

\usepackage[numbers]{natbib}                      
\usepackage[preprint]{neurips_2025}       

\usepackage{graphicx}%

\usepackage{multirow}%
\usepackage{subcaption}
\usepackage{amsmath,amssymb,amsfonts}%
\usepackage{amsthm}%
\usepackage{mathrsfs}%
\usepackage[title]{appendix}%
\usepackage{xcolor}%
\usepackage{textcomp}%
\usepackage{manyfoot}%
\definecolor{DarkGreen}{HTML}{013220}

\usepackage{booktabs}%
\usepackage{algorithm}%
\usepackage{algorithmicx}%
\usepackage{algpseudocode}%
\usepackage{listings}%
\usepackage{longtable}
\usepackage[utf8]{inputenc} 
\usepackage[T1]{fontenc}    
\usepackage{url}            
\usepackage{booktabs}       
\usepackage{amsfonts}       
\usepackage{nicefrac}       
\usepackage{microtype} 
\usepackage[export]{adjustbox}    
\usepackage[breaklinks=true]{hyperref}

\makeatletter
\g@addto@macro{\UrlBreaks}{\do\-}
\@submissionfalse

\title{How Hungry is AI? Benchmarking Energy, Water, and Carbon Footprint of LLM Inference}

\author{
  Nidhal Jegham\(^1,^2\) \\
  \texttt{nidhal.jegham@uri.edu}
  \And
  Marwan Abdelatti\(^3\) \\
  \texttt{mabdelat@providence.edu}
  \And
  Chan Young Koh\(^1\) \\
  \texttt{ckoh04@uri.edu}
  \And
  Lassad Elmoubarki\(^2\) \\
  \texttt{lassad.elmoubarki@tbs.rnu.tn}
  \And
  Abdeltawab Hendawi\(^1\)\thanks{Corresponding author.} \\
  \texttt{hendawi@uri.edu}
}

\begin{document}

\maketitle
\vspace{-2em}   

\begin{center}
\(^1\) University of Rhode Island
\quad
\(^2\) University of Tunis
\quad
\(^3\) Providence College
\\[1ex]
Live Dashboard: 
\href{https://app.powerbi.com/view?r=eyJrIjoiZjVmOTI0MmMtY2U2Mi00ZTE2LTk2MGYtY2ZjNDMzODZkMjlmIiwidCI6IjQyNmQyYThkLTljY2QtNDI1NS04OTNkLTA2ODZhMzJjMTY4ZCIsImMiOjF9}
{\textcolor[HTML]{013220}{\textbf{Power BI Dashboard}}}
\end{center}

\begin{abstract}

This paper introduces an infrastructure-aware benchmarking framework for quantifying the environmental footprint of LLM inference across 30 state-of-the-art models in commercial datacenters. The framework combines public API performance data with company-specific environmental multipliers and statistical inference of hardware configurations. We additionally utilize cross-efficiency Data Envelopment Analysis (DEA) to rank models by performance relative to environmental cost and provide a dynamically updated dashboard that visualizes model-level energy, water, and carbon metrics. Results show the most energy-intensive models exceed 29~Wh per long prompt, over 65$\times$ the most efficient systems. Even a 0.42~Wh short query, when scaled to 700M queries/day, aggregates to annual electricity comparable to 35{,}000 U.S. homes, evaporative freshwater equal to the annual drinking needs of 1.2M people, and carbon emissions requiring a Chicago-sized forest to offset. These findings highlight a growing paradox: as AI becomes cheaper and faster, global adoption drives disproportionate resource consumption. Our methodology offers a standardized, empirically grounded basis for sustainability benchmarking and accountability in AI deployment.


\end{abstract}

\section{Introduction}

Large language models (LLMs) have moved beyond research labs and are now embedded in search engines, virtual assistants, education platforms, and enterprise tools \cite{google2023search, qin2023toolllm, hannan2023ai, rajpurkar2023ai}. Models like GPT-4o \cite{openai_gpt4o} and Claude-3.7 Sonnet \cite{anthropic_claude} represent state-of-the-art systems, while open-source alternatives such as LLaMA-3 \cite{meta_llama3} and DeepSeek-V3 \cite{deepseek2024} reflect growing accessibility and experimentation. On top of that, the emergence of reasoning models such as DeepSeek-R1 \cite{guo2025deepseek}, o1 \cite{openai_o1}, and o3-mini \cite{openai_o3} marks a shift toward multi-step logic and chain-of-thought reasoning.

However, the advancement of LLMs does involve shortcomings in environmental aspects. Training GPT-3 is estimated to consume 1,287 megawatt-hours (MWh) of electricity and emit over 550 metric tons of CO\textsubscript{2} equivalent ($\text{CO}_2\text{e}$) \cite{patterson2021carbon}, while requiring more than 700 kiloliters (kL) of water for cooling alone \cite{li2023making}, enough to fill a quarter of an Olympic-sized swimming pool. Yet while training has been the focus of sustainability discussions, inference is emerging as the primary contributor to environmental costs. In contrast to training, which is conducted once or at intervals, inference occurs consistently and on a large scale. Recent estimates suggest inference can account for up to 90\% of a model's total lifecycle energy use \cite{yang2024trends, lacoste2022codecarbon}.

Despite the growing environmental footprint of large-scale model deployment, a standard method to quantify the cost of inference at the prompt level remains absent. A core obstacle to developing more accurate assessments is the lack of model-specific inference data for commercial AI models. Existing environmental reports tend to aggregate emissions across entire cloud infrastructures without disaggregating by model or workload \cite{microsoft2024environmental, google2024environmental}. This lack of public information hinders independent verification and undermines both scientific benchmarking and policy efforts aimed at regulating AI’s true environmental cost.

To address these issues, we introduce a novel infrastructure-aware benchmarking framework to quantify the operational environmental footprint of LLM inference at the per-prompt level as deployed in data centers. Unlike existing studies \cite{li2023making, lacoste2022codecarbon, husom2024melodi}, our method adopts a more comprehensive strategy by integrating performance metrics such as latency and throughput from public APIs with published GPU and system power specifications. Furthermore, we scale these combined data points using company-specific multipliers, including Power Usage Effectiveness (PUE) \cite{GreenGridPUE, ISOIEC30134-2}, Water Usage Effectiveness (WUE) \cite{GreenGridPUE, ISOIEC30134-2}, and Carbon Intensity Factors (CIF) \cite{EPAeGRID, IEAEmissionFactors} to account for infrastructural overhead. This method enables us to evaluate the energy, water, and carbon effects of both open-source and proprietary models, a gap that, to our knowledge, has not been comprehensively explored in prior research. Additionally, we employ statistical analysis, including ANOVA and Tukey HSD, to estimate underlying hardware configurations. To enhance transparency and reproducibility, we also developed an automated and interactive Power BI dashboard that visualizes the daily fluctuations in the energy, water, and carbon footprint of an extended list of models across multiple data centers. This novel dashboard incorporates new models as they get released. Moreover, to contextualize resource use relative to model capability, we apply cross-efficiency Data Envelopment Analysis (DEA) to assess how effectively each model converts environmental inputs into performance. As a key application of this framework, we perform a case study to estimate the footprint of GPT-4o text generation based on scaled usage data. We further extend our analysis to GPT-5, focusing on the disparities in energy consumption between queries that involve different levels of reasoning. Our framework enables infrastructure-aware decision-making, empowers accountability, and provides a foundational step toward sustainability standards in AI deployment.

The remainder of the paper is organized as follows. Section~\ref{sec:related_work} reviews existing studies on the environmental impact of LLMs. Section~\ref{sec:preliminaries} introduces key concepts, including hardware configurations and environmental multipliers. Section~\ref{sec:methodology} details our framework for estimating inference-phase cost. Section~\ref{sec:results} presents findings across 30 models. Section \ref{section: Appendix E} provides a focused analysis of GPT-4o’s annual environmental footprint and section \ref{section: gpt-5} analyzes the impact of GPT-5's adapative model routing. Section~\ref{sec:discussion} outlines key insights and implications. Section~\ref{sec:conclusion} summarizes the main takeaways and limitations and directions for future work.

\section{Related Work}  \label{sec:related_work}

The environmental impact of AI systems has garnered increasing attention in recent years, with a growing body of work attempting to quantify the energy, carbon, and water costs associated with training and deploying LLMs.

Li et al. \cite{li2023making} analyzed GPT-3’s freshwater consumption, estimating over 5 million liters used during training and projecting that AI-related withdrawals could reach 6.6 trillion liters annually by 2027. Although their spatiotemporal methodology is a significant early contribution, it overlooks carbon emissions, depends on an outdated model, and requires previous knowledge of energy usage, which restricts its scalability. In parallel, Strubell et al. \cite{strubell2020energy} estimated carbon emissions from training BERT and GPT-2 by accounting for GPU, CPU, and DRAM power draw alongside PUE adjustments. However, their analysis excludes inference and infrastructural overhead. Similar limitations appear in Meta’s LLaMA reports \cite{meta_llama3,llama1, llama2}, which provide carbon footprints based on GPUs' TDPs but disregard water use, system-wide energy consumption, and the inference phase entirely.

Regarding inference, Husom et al. \cite{husom2024melodi} (MELODI) measure real-time energy consumption of GPUs and CPUs at the prompt level, but they neglect carbon emissions, water usage, and infrastructure overhead, only concentrating on small-scale open-source models. Samsi et al. \cite{samsi2023watts} measure GPU power draw across prompt lengths but exclude proprietary systems and broader environmental factors, lacking a standardized scaling method for production-level inference. Yang et al. \cite{yang2024double} evaluate over 1,200 vision models and introduce an energy-efficiency score. However, their analysis does not include LLMs, API-based deployments, or essential infrastructure considerations like PUE and WUE.

Complementary studies, including Luccioni et al. \cite{luccioni2024power}, assess general-purpose and task-specific models in the A100 systems. While they provide valuable cross-model insights, they do not consider proprietary models, water usage, or carbon emissions. CodeCarbon \cite{lacoste2022codecarbon} calculates carbon footprints based on device-level data and regional carbon intensity, but it lacks the granularity needed for prompt-level analysis and does not work with API-based inferences. On a larger scale, Harding et al. \cite{harding2024watts} connect AI adoption to national productivity, allowing for extrapolation of energy and carbon effects. Though this provides a useful overarching view, it overlooks variability in per-prompt inference, the behavior of specific models, and the infrastructure used for deployment.

Most efforts focus on training and local model evaluation, lacking standardized, scalable methods, ignoring infrastructural overhead, and omitting resource categories such as water consumption and carbon emissions. Our work addresses these gaps by integrating API-based performance metrics with GPU and system power specifications and environmental multipliers to estimate the environmental impact of LLM inference at the prompt level in data centers. We infer deployment infrastructure through statistical analysis and apply DEA to contextualize environmental impact versus performance. Additionally, we conduct two case studies estimating GPT-4o's annual environmental footprint based on scaled usage data and analyzing the impact of GPT-5's adapative model routing, providing the first infrastructure-aware, prompt-level benchmark of inference sustainability at scale.

\section{Preliminaries} \label{sec:preliminaries}

To capture infrastructure-level overhead in data center operations, we apply three standard environmental multipliers: Power Usage Effectiveness (PUE) \cite{GreenGridPUE, ISOIEC30134-2}, Water Usage Effectiveness (WUE) \cite{GreenGridPUE, ISOIEC30134-2}, and Carbon Intensity Factor (CIF) \cite{EPAeGRID, IEAEmissionFactors}. 

\textbf{PUE} accounts for non-computational energy overheads such as cooling, lighting, and power distribution. Defined as the ratio of total data center energy consumption to IT-specific energy use. 
 
\textbf{WUE} captures the water used per kilowatt-hour of IT energy, encompassing on-site cooling (Scope 1), off-site electricity generation (Scope 2), and embodied water from hardware manufacturing and transport (Scope 3). WUE can be computed based on either water withdrawal (the total volume drawn from natural or municipal sources) or water consumption (the portion of withdrawn water permanently lost, primarily through evaporation). 

\textbf{CIF} measures carbon emissions per kilowatt-hour of energy consumed, largely driven by the regional electricity mix. Emissions are categorized as direct on-site combustion (Scope 1), off-site electricity generation (Scope 2), and embodied emissions from manufacturing and transport (Scope 3).

\section{Methodology} \label{sec:methodology}

This section presents our novel methodology for estimating the environmental footprint of LLM inference. Our framework integrates model-specific performance metrics with infrastructure-level environmental multipliers to calculate operational energy consumption, water usage, and carbon emissions per query. We also evaluate eco-efficiency using DEA, mapping sustainability trade-offs against a composite performance benchmark, and develop an interactive dashboard for a more thorough analysis.

\subsection{Model Selection and Hardware Estimation}

\begin{table}[]
\caption{Deployment and infrastructure specifications of models.}
\label{tab:model-specs}
\vspace{0.5em}

\resizebox{\textwidth}{!}{\begin{tabular}{cccccccccc}
\hline
\textbf{Model} & 
\textbf{\begin{tabular}[c]{@{}c@{}}Launch \\ Date\end{tabular}} & 
\textbf{Company} & 
\textbf{Host} & 
\textbf{Hardware} & 
\textbf{\begin{tabular}[c]{@{}c@{}}Critical \\ Power \\ (kW)\end{tabular}} & 
\textbf{\begin{tabular}[c]{@{}c@{}}PUE \\ \end{tabular}} & 
\textbf{\begin{tabular}[c]{@{}c@{}}WUE \\ (on-site, L/kWh)\end{tabular}} & 
\textbf{\begin{tabular}[c]{@{}c@{}}WUE \\ (off-site, L/kWh)\end{tabular}} & 
\begin{tabular}[c]{@{}c@{}}\textbf{CIF} \\ \textbf{(kgCO\textsubscript{2}e/kWh)}\end{tabular}
\\ \hline
GPT-4.1              & Apr, 2025                                                       & \multirow{11}{*}{OpenAI}   & \multirow{11}{*}{Microsoft Azure} & \multirow{11}{*}{DGX H200/H100 \cite{grimm2024dgxh200, nvidia2023hopper}} & \multirow{11}{*}{10.20 \cite{idlepower}} & \multirow{11}{*}{1.12 \cite{walsh2022azure}} & \multirow{11}{*}{0.30 \cite{solomon2024sustainable}} & \multirow{11}{*}{4.35 \cite{wri2024wateruse}} & \multirow{11}{*}{0.35 \cite{microsoft2024sustainability}} \\
GPT-4.1 mini         & Apr, 2025                                                       &                            &                                   &                                 &                                              &                                         &                                               &                                              &                          \\
GPT-4.1 nano         & Apr, 2025                                                       &                            &                                   &                                 &                                              &                                         &                                               &                                              &                          \\
o4-mini (high)       & Apr, 2025                                                       &                            &                                   &                                 &                                              &                                         &                                               &                                              &                          \\

o3                   & Apr, 2025                                                       &                            &                                   &                                 &                                              &                                         &                                               &                                              &                          \\
o3-mini (high)       & Jan, 2025                                                       &                            &                                   &                                 &                                              &                                         &                                               &                                              &                          \\
o3-mini              & Jan, 2025                                                       &                            &                                   &                                 &                                              &                                         &                                               &                                              &                          \\
o1                   & Dec, 2024                                                       &                            &                                   &                                 &                                              &                                         &                                               &                                              &                          \\
o1-mini              & Sep, 2024                                                       &                            &                                   &                                 &                                              &                                         &                                               &                                              &                          \\
GPT-4o (Mar '25)     & May, 2024                                                       &                            &                                   &                                 &                                              &                                         &                                               &                                              &                          \\ \hline
GPT-4o mini          & July, 2024                                                      & \multirow{3}{*}{OpenAI}    & \multirow{3}{*}{Microsoft Azure}  & \multirow{3}{*}{DGX A100\textsuperscript{*}}       & \multirow{3}{*}{6.50\cite{nvidia_dgx_a100_2020}} & \multirow{3}{*}{1.12} & \multirow{3}{*}{0.30} & \multirow{3}{*}{4.35} & \multirow{3}{*}{0.35} \\
GPT-4 Turbo          & Nov, 2023                                                       &                            &                                   &                                 &                                              &           &          &             &                          \\
GPT-4                & Mar, 2023                                                       &                            &                                   &                                 &                                              &           &          &             &                          \\ \hline
DeepSeek-R1          & Jan, 2025                                                       & \multirow{2}{*}{Deepseek}  & \multirow{2}{*}{Deepseek}         & \multirow{2}{*}{DGX H800 \cite{deepseek2024}} & \multirow{2}{*}{10.20 \cite{nvidia_dgx_h800}} & \multirow{2}{*}{1.27 \cite{wu2024datacenter}}  & \multirow{2}{*}{1.20 \cite{wu2024datacenter}} & \multirow{2}{*}{6.016 \cite{wri2024wateruse}} & \multirow{2}{*}{0.6 \cite{smith2023renewable}} \\
DeepSeek-V3          & Dec, 2024                                                       &                            &                                   &                                 &                                              &           &          &             &                          \\ \hline
DeepSeek-R1          & Jan, 2025                                                       & \multirow{2}{*}{Deepseek}  & \multirow{2}{*}{Microsoft Azure}         & \multirow{2}{*}{DGX H200/H100} & \multirow{2}{*}{10.20} & \multirow{2}{*}{1.12}  & \multirow{2}{*}{0.30} & \multirow{2}{*}{4.35} & \multirow{2}{*}{0.35} \\
DeepSeek-V3          & Dec, 2024                                                       &                            &                                   &                                 &                                              &           &          &             &                          \\ \hline
Claude-3.7 Sonnet    & Feb, 2025                                                       & \multirow{4}{*}{Anthropic} & \multirow{4}{*}{AWS}              & \multirow{4}{*}{DGX H200/H100 \cite{aws2023p5, aws2024h200}}  & \multirow{4}{*}{10.20} & \multirow{4}{*}{1.14 \cite{amazon2023sustainability}}  & \multirow{4}{*}{0.18 \cite{amazon2023sustainability}} & \multirow{4}{*}{5.11 \cite{wri2024wateruse}} & \multirow{4}{*}{0.287 \cite{ElectricityMaps2025}} \\
Claude-3.5 Sonnet    & Jun, 2024                                                       &                            &                                   &                                 &                         &           &          &             &                          \\
Claude-3.5 Haiku     & Nov, 2024                                                       &                            &                                   &                                 &                         &           &          &             &                          \\ \hline
LLaMA-3.3 70B        & Dec, 2024                                                       & \multirow{10}{*}{Meta}     & \multirow{10}{*}{AWS}             & \multirow{10}{*}{DGX H200/H100} & \multirow{10}{*}{10.20} & \multirow{10}{*}{1.14} & \multirow{10}{*}{0.18} & \multirow{10}{*}{5.11} & \multirow{10}{*}{0.287} \\
LLaMA-3.2-vision 90B & Sep, 2024                                                       &                            &                                   &                                 &                         &           &          &             &                          \\
LLaMA-3.2-vision 11B & Sep, 2024                                                       &                            &                                   &                                 &                         &           &          &             &                          \\
LLaMA-3.2 3B         & Sep, 2024                                                       &                            &                                   &                                 &                         &           &          &             &                          \\
LLaMA-3.2 1B         & Sep, 2024                                                       &                            &                                   &                                 &                         &           &          &             &                          \\
LLaMA-3.1-405B       & Jul, 2024                                                       &                            &                                   &                                 &                         &           &          &             &                          \\
LLaMA-3.1-70B        & Jul, 2024                                                       &                            &                                   &                                 &                         &           &          &             &                          \\
LLaMA-3.1-8B         & Jul, 2024                                                       &                            &                                   &                                 &                         &           &          &             &                          \\
LLaMA-3-70B          & Apr, 2024                                                       &                            &                                   &                                 &                         &           &          &             &                          \\
LLaMA-3-8B           & Apr, 2024                                                       &                            &                                   &                                 &                         &           &          &             &                          \\ \hline
\end{tabular}}
\vspace{-0.5em}
\begin{flushleft}
\tiny{\textsuperscript{*}DGX A100 was estimated for GPT-4o mini, GPT-4 Turbo, and GPT-4. Justification and estimation details are provided in Section~\ref{sec:hardware_estimation}.} \newline

\end{flushleft}
\end{table}

We analyze 30 large language models across OpenAI, Anthropic, Meta, and DeepSeek. Table \ref{tab:model-specs} summarizes each model’s deployment context, including provider, cloud host, hardware type and specifications, and company-specific environmental multipliers (PUE, WUE, CIF). All models are usually run on NVIDIA DGX systems using A100, H100, H200, or H800 GPUs \cite{grimm2024dgxh200, nvidia_dgx, shehabi2024datacenter, borkar2024microsoft, nvidia_project_ceiba}. U.S.-based providers such as OpenAI and Anthropic have acquired large volumes of H200 and H100 chips \cite{nvidia2023hopper, aws2023p5, aws2024h200}, making them the most probable choice for recent deployments. DeepSeek, which operates under U.S. export restrictions, uses the H800, NVIDIA’s export-compliant GPU for the Chinese market \cite{nvidia_dgx_h800, nytimes2025}. Both the H200 and H800 retain the same Hopper architecture and peak power draw as the H100, with system-level energy characteristics that are nearly identical \cite{nvidia2025dgx}. While the H200 achieves greater energy efficiency due to faster memory and higher bandwidth, and the H800 may exhibit reduced performance due to export-related firmware limitations, both maintain the same peak power draw, thermal design profile, and system-level utilization characteristics as the H100 \cite{nvidia_dgx_h800, nvidia2025dgx}. These architectural differences affect throughput and latency, resulting in higher or lower energy consumed per token, but do not impact total system power demand under load. We therefore treat H100, H200, and H800 as equivalent in our power modeling, since our estimates are based on power draw and utilization rather than task-level performance.

Environmental multipliers such as PUE, WUE, and CIF are assigned according to each cloud provider’s data center locations and corresponding regional grid characteristics. For OpenAI and DeepSeek models hosted on Microsoft Azure, we use Azure-reported PUE and site-level WUE values, while CIF and source-level WUE are derived from the specific geographic locations of Microsoft data centers around the world. For AWS-hosted models, including those from Anthropic and Meta, we apply AWS-reported PUE and site-level WUE, and compute CIF and source-level WUE based on the regional distribution of AWS data centers used for inference. For DeepSeek models that are deployed in Chinese datacenters, we adopt the average PUE and site-level WUE of the thirty most efficient data centers in China, while CIF and source-level WUE are determined using the regional locations of its known or reported data center deployments.

\subsection{Per-Query Energy Consumption Estimation}
To quantify the energy required for a single inference, we introduce a probabilistic framework that captures the stochastic nature of LLM workloads. The model integrates standardized performance data \cite{artificialanalysis2025}, which report latency to first-token generation (\(L\)) and tokens-per-second (TPS, denoted \(R\)) across empirical quantiles (5th, 25th, 50th, 75th, and 95th percentiles) and three representative prompt configurations: short-form (100 input, 300 output tokens), medium (1,000 input, 1,000 output), and long-form (10,000 input, 1,500 output), reflecting variability across multiple test runs for each model and prompt configuration.

To model realistic runtime behavior, we construct a joint distribution of \(L\) and \(R\) using a Gaussian copula with correlation coefficient \(\rho = -0.3\), capturing the negative dependence typically observed between latency and TPS. From this distribution, we draw 10,000 correlated samples \((L_i, R_i)\), each representing one plausible inference scenario. The culmination of this infrastructure-aware framework is the introduction of our novel
 formula to precisely estimate the per-query energy consumption:

Let \(L_i\) captures the initialization latency and \(\tfrac{\text{Output Length}}{R_i}\) represents the time it takes to generate the response. Also, let \(P_{\text{GPU}}\) and \(P_{\text{non-GPU}}\) denote the rated power draw (in kW) of the GPU subsystem and the non-GPU subsystem (e.g., CPUs, SSDs, network, and cooling control electronics), respectively. The parameters \(U_{\text{GPU,min}}\) and \(U_{\text{GPU,max}}\) represent the minimum and maximum GPU utilization fractions observed during inference, while \(U_{\text{non-GPU}}\) represents the average utilization fraction for non-GPU components. PUE factor is also incorporated to account for datacenter-level overheads.

We compute energy consumption at the lower and upper utilization bounds as:
\begin{equation} \label{eq:eq1}
E_{i,\{ \min,\max \}} = 
\underbrace{\left(\frac{L_i + \tfrac{\text{Output Length}}{R_i}}{3600}\right)}_{\text{Total inference time } (T_i, \text{ hours})}
\times 
\left[
    \underbrace{P_{\text{GPU}} \times U_{\text{GPU,\{min,max\}}}}_{\text{GPU power (kW)}} 
    + 
    \underbrace{P_{\text{non-GPU}} \times U_{\text{non-GPU}}}_{\text{Non-GPU power (kW)}}
\right]
\times \text{PUE}
\end{equation}

We also define an expected per-query energy as a weighted combination of both scenarios ($w_{\max}=0.5$), and the framework aggregates all Monte Carlo draws to produce a distribution of per-query energy outcomes. The final metrics are reported as the sample mean and standard deviation:

\begin{equation}\label{eq:eq3_4}
E_{i,\text{exp}} = w_{\max} E_{i,\max} + (1 - w_{\max}) E_{i,\min}, \quad 
\bar{E}_{\text{query}} = \mathbb{E}[E_{i,\text{exp}}], \quad 
\sigma_{E_{\text{query}}} = \sqrt{\text{Var}[E_{i,\text{exp}}]}
\end{equation}

This stochastic formulation captures variability in runtime, hardware utilization, and data-center efficiency, enabling robust and reproducible estimation of per-query energy consumption across diverse inference conditions.


\subsection{Hardware-Class Attribution}
We stratify LLMs into five hardware classes based on model size: \textbf{Nano} (\textless{}7B), \textbf{Micro} (7--20B), \textbf{Small} (20--40B), \textbf{Medium} (40--70B), and \textbf{Large} (>70B), assigning 1, 2, 4, or 8 GPUs accordingly. Models that do not disclose parameter counts, such as OpenAI and Anthropic flagship models (e.g., GPT-4o, Claude-3.7 Sonnet), are classified as \textbf{Large}, OpenAI Mini variants (e.g., GPT-4o mini) as \textbf{Medium}, and models labeled ``Nano'' such as GPT-4.1 nano as \textbf{Small} based on reported model performance (e.g., TPS, latency, and reasoning capabilities) \cite{artificialanalysis2025}. 

AI companies and cloud providers typically rely on dynamic batching to optimize GPU utilization while maintaining low latency \cite{nvidia_triton_batcher_2024}. Although actual batch sizes fluctuate depending on incoming demand, they are generally constrained to a narrow range below 16 to preserve responsiveness. Benchmarks \cite{artificialanalysis2025} show that even for large prompts, most models maintain a first-token latency below one second. Moreover, prior studies \cite{llminfbench2024, splitwise2024} show that these latency values are consistent with batch sizes in the range of 4 to 16. This suggests that real-world deployments prioritize small, latency-sensitive batches over maximal throughput. Accordingly, we adopt a batch size of 8 for all primary calculations, as it represents a practical midpoint between common deployment scenarios. A detailed sensitivity analysis exploring the impact of alternative batch sizes is provided in Appendix \ref{sec: Appendix B}. The number of GPUs and their allocated power draw utilization rates for H100 systems are estimated from Splitwise \cite{splitwise2024}, the Latency Processing Unit study \cite{lpu2024}, and LLM-Inference-Bench \cite{llminfbench2024}. For A100 systems, we adopt measurements from Patel et al. and Kakolyris et al.'s work \cite{patel2024characterizing, kakolyris2024slo}. Per-request GPU and non-GPU utilization rates are calculated as:
\begin{equation}
U_{\text{GPU total}} = \frac{G \times D_{\text{GPU}}}{N \times B}, \quad\quad 
U_{\text{non-GPU total}} = \frac{G \times D_{\text{non-GPU}}}{N \times B}
\end{equation}

where \(G\) is the number of GPUs assigned per model, \(N = 8\) is the number of GPUs per node, and \(B = 8\) is the batch size. \(D_{\text{GPU}}\) denotes the assigned GPUs' power draw, expressed as a fraction of their maximum power draw, while \(D_{\text{non-GPU}} = 0.5\) represents the conservatively assigned fixed utilization fraction for non-GPU components (e.g., CPU, memory, storage, cooling), relative to their peak power draw~\cite{idlepower}. We exclude idle power consumption from unutilized GPUs in partially loaded nodes, as deployment-specific telemetry is unavailable to determine whether such capacity is reassigned, load-balanced, or remains idle. Table~\ref{tab:energy-share} summarizes GPU and non-GPU power utilization rates across model classes. Values are rounded to typical intervals observed during inference, accounting for input processing spikes, output length, decoding complexity, and a batch size of 8 parallel requests.

\begin{table}[h]
  \centering
\caption{Estimated node-level GPU and non-GPU utilization by model class for H100 and A100.}

  \label{tab:energy-share}
\begin{tabular}{lcccccl}
\hline
\textbf{Class}                     & \begin{tabular}[c]{@{}c@{}}\text{\textbf{GPU}} \\ \textbf{Count}\end{tabular} & \begin{tabular}[c]{@{}c@{}}\textbf{D\textsubscript{GPU}} \\ \textbf{(H100)}\end{tabular} & \begin{tabular}[c]{@{}c@{}}\textbf{D\textsubscript{GPU}}  \\\textbf{(A100)}\end{tabular} & \begin{tabular}[c]{@{}c@{}}\textbf{U\textsubscript{GPU total}} \\ \textbf{(H100)}\end{tabular} & \begin{tabular}[c]{@{}c@{}}\textbf{U\textsubscript{GPU total}} \\ \textbf{(A100)}\end{tabular} & \textbf{U\textsubscript{non-GPU total}} \\ \hline
Nano          & 1        & 35--65\%                                                                    & 80--90\%                                                       & 0.55--1.00\%                                                             & 1.25--1.5\%                                                              & 0.87\%                     \\
Micro              & 1        & 50--80\%                                                                    & 90--100\%                                                      & 0.75--1.25\%                                                              & 1.5--1.6\%                                                              & 0.87\%                     \\
Small            & 2        & 55--80\%                                                                    & N/A                                                            & 1.70--2.50\%                                                                & N/A                                                                      & 1.6\%                      \\
Medium           & 4        & 50--70\%                                                                    & 100--110\%                                                     & 3.00--4.50\%                                                             & 6.25--7\%                                                                & 3.125\%                    \\
Large  & 8        & 45--60\%                                                                    & 100--120\%                                                     & 5.50--7.50\%                                                              & 12.5--15.0\%                                                             & 6.25\%                     \\ \hline
\end{tabular}
\end{table}

\subsubsection{GPT-4, GPT-4 Turbo, and GPT-4o mini Hardware Estimation} \label{sec:hardware_estimation}

\begin{figure}
    \centering
    \includegraphics[width=1\linewidth]{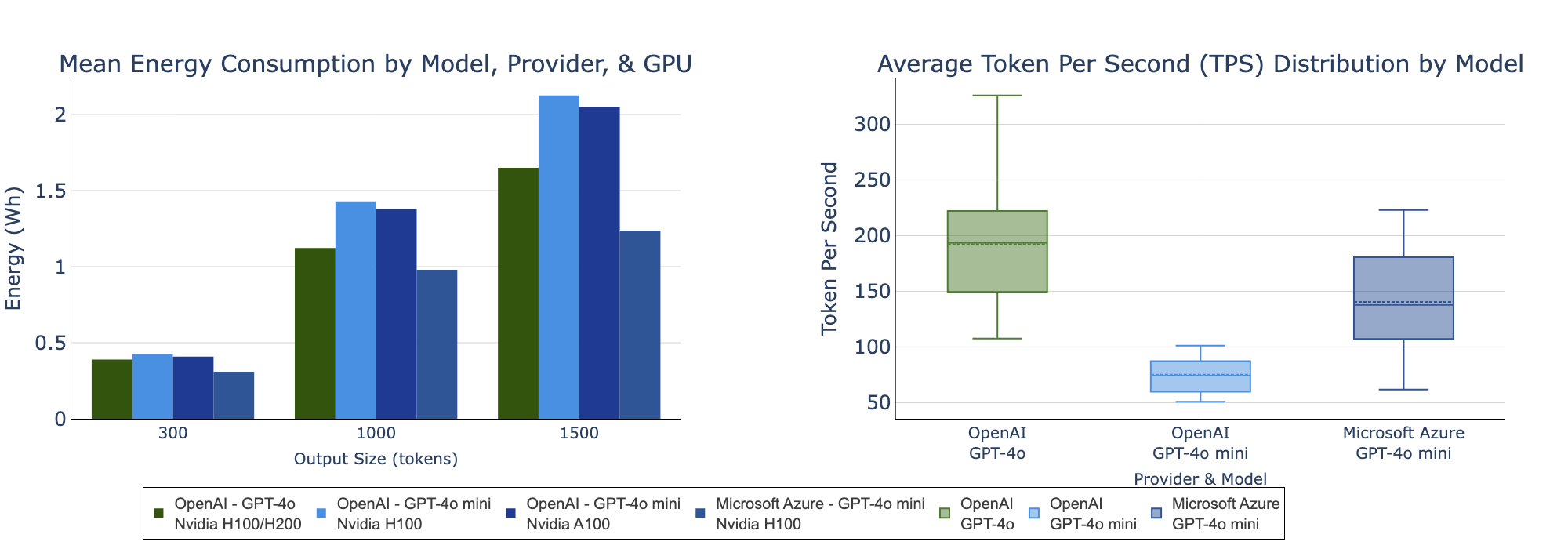}
\caption{(Left) Mean energy consumption of GPT-4o and GPT-4o mini across providers and GPU types, measured by output size. (Right) Distribution of TPS (averaged across output sizes)}

    \label{fig:GPT4oAnalysis}
    
\end{figure}

In our experiment, we observed a performance discrepancy: GPT-4o mini showed significantly lower throughput and higher latency on OpenAI’s API compared to Microsoft Azure under identical prompt settings, as shown in Figure \ref{fig:GPT4oAnalysis}. Both variants also underperformed relative to OpenAI’s GPT-4o, with 60\% and 27\% lower TPS, respectively. Given GPT-4o mini’s smaller size and H200’s architectural advantages, its performance would be expected to match or exceed GPT-4o if served on H200 infrastructure. The observed gap is inconsistent with H200 deployment and suggests that GPT-4o mini is running on A100 or H100 systems. Notably, Azure’s version outperforms OpenAI’s by 47\% on average, further supporting the likelihood that Azure uses H100 and OpenAI retains A100. Therefore, to validate our hardware estimations, we tested this hypothesis using two-way ANOVA and Tukey HSD (Table \ref{tab:tukey-pvals}). At 300-token prompts, energy consumption was statistically similar across platforms, as expected given the small computational load. However, at larger output sizes, significant differences emerged: OpenAI’s presumed A100 deployment differed from Azure’s H100 deployment with $p < 0.05$, and Azure’s H100 also outperformed OpenAI’s assumed H100 with $p < 0.05$, reinforcing the likelihood that OpenAI’s GPT-4o mini is not served on H100. We therefore consider GPT-4o mini to be running on A100. Additionally, with reports that GPT-4 was trained and deployed on A100 systems \cite{patel2023gpt4}, and given the architectural continuity between GPT-4 and GPT-4 Turbo and their low throughput, high latency, and impending deprecation \cite{openai_deprecations}, we also consider they are running on A100 architecture since it is unlikely that they have migrated to newer hardware.

\begin{table}[ht]
\caption{Tukey HSD Adjusted $p$-values for energy consumption differences by provider, GPU system, and prompt size}
\label{tab:tukey-pvals}
\centering
\begin{tabular}{llccc}
\toprule
\textbf{Group 1} & \textbf{Group 2} & \textbf{300 tokens} & \textbf{1000 tokens} & \textbf{1500 tokens} \\
\midrule
Azure (H100) & OpenAI (A100) & 0.979 & 0.0009 & $<$0.0001 \\
Azure (H100) & OpenAI (H100) & 0.951 & 0.0001 & $<$0.0001 \\
\bottomrule
\end{tabular}
\vspace{0.5em}

\end{table}

\subsection{Per-Query Water Consumption and Carbon Emissions Estimation}

This study focuses exclusively on operational emissions and resource consumption during the inference phase of the model. Accordingly, embodied emissions and water use from hardware manufacturing and supply chains (Scope 3) are excluded due to their limited relevance to real-time deployment and the risk of inflating per-query estimates when applied without deployment-specific attribution or when model lifecycles remain ongoing. For water usage, we focus solely on water consumption (water permanently removed from the source). For carbon emissions, we exclude Scope 1 emissions as they are generally negligible compared to Scope 2 emissions due to the infrequent use of on-site fuel combustion for backup generators and facility heating in data centers \cite{aslan2025toward}. For example, Scope 1 emissions accounted for only 1.6\% of Microsoft’s Scope 2 emissions in 2023 \cite{microsoft2024sustainability}, a figure that includes executive air travel, ground transportation, refrigerant leakage, and on-site fuel use, further diminishing the share attributable to data center operations. Accordingly, our analysis focuses exclusively on Scope 2 emissions, which capture the carbon intensity of electricity consumed during inference. A more detailed discussion of these considerations is provided in Appendix \ref{sec: Appendix A}.

Water consumption and carbon emissions per query are calculated as:

\begin{equation}\label{eq:water_cons}
\text{Water (L)} = 
\underbrace{\frac{E_{\text{query}}}{\text{PUE}} \cdot \text{WUE}_{\text{site}}}_{\text{On-site cooling}} 
+ 
\underbrace{E_{\text{query}} \cdot \text{WUE}_{\text{source}}}_{\text{Off-site electricity}} \quad\quad 
\end{equation}

\vspace{-0.8em}
\begin{equation}\label{eq:carbon_emiss}
\text{Carbon (kgCO}_2\text{e)} = E_{\text{query}} \cdot \text{CIF}
\end{equation}

\subsection{Eco-Efficiency via Data Envelopment Analysis (DEA)}

We apply cross-efficiency DEA to evaluate the effectiveness of each model in converting environmental resources into functional intelligence. Inputs include per-query energy consumption, PUE, $\text{WUE}_{\text{source}}$, $\text{WUE}_{\text{site}}$, and CIF. The output is the Artificial Intelligence Index, a composite score weighted across multiple benchmark domains \cite{artificialanalysis2025}. Specifically, reasoning and knowledge tasks (MMLU-Pro~\cite{mmlu_pro}, HLE~\cite{hle2025}, GPQA~\cite{gpqa2023}) collectively contribute 50\% of the index (1/6 each); mathematical proficiency (MATH-500~\cite{math500}, AIME~\cite{aime2024}) contributes 25\% (1/8 each); and coding ability (SciCode~\cite{scicode2024}, LiveCodeBench~\cite{livecodebench2024}) accounts for the remaining 25\% (1/8 each).
 
In contrast to standard Charnes-Cooper-Rhodes (CCR) or Banker-Charnes-Cooper (BCC) models, which enable each model to choose its optimal weightings, sometimes inflating performance, cross-efficiency assesses each model based on its own and all peer weightings. This approach reduces self-evaluation bias and recognizes models that maintain strong performance from various efficiency viewpoints. The resulting scores offer a more robust and comparative measure of eco-efficiency. Full results and additional discussion are provided in Appendix~\ref{sec: Appendix C}.

\subsection{Power BI Dashboard}
\begin{figure}[h!]
    \centering
    \begin{subfigure}[t]{0.48\linewidth}
        \centering
        \includegraphics[width=\linewidth]{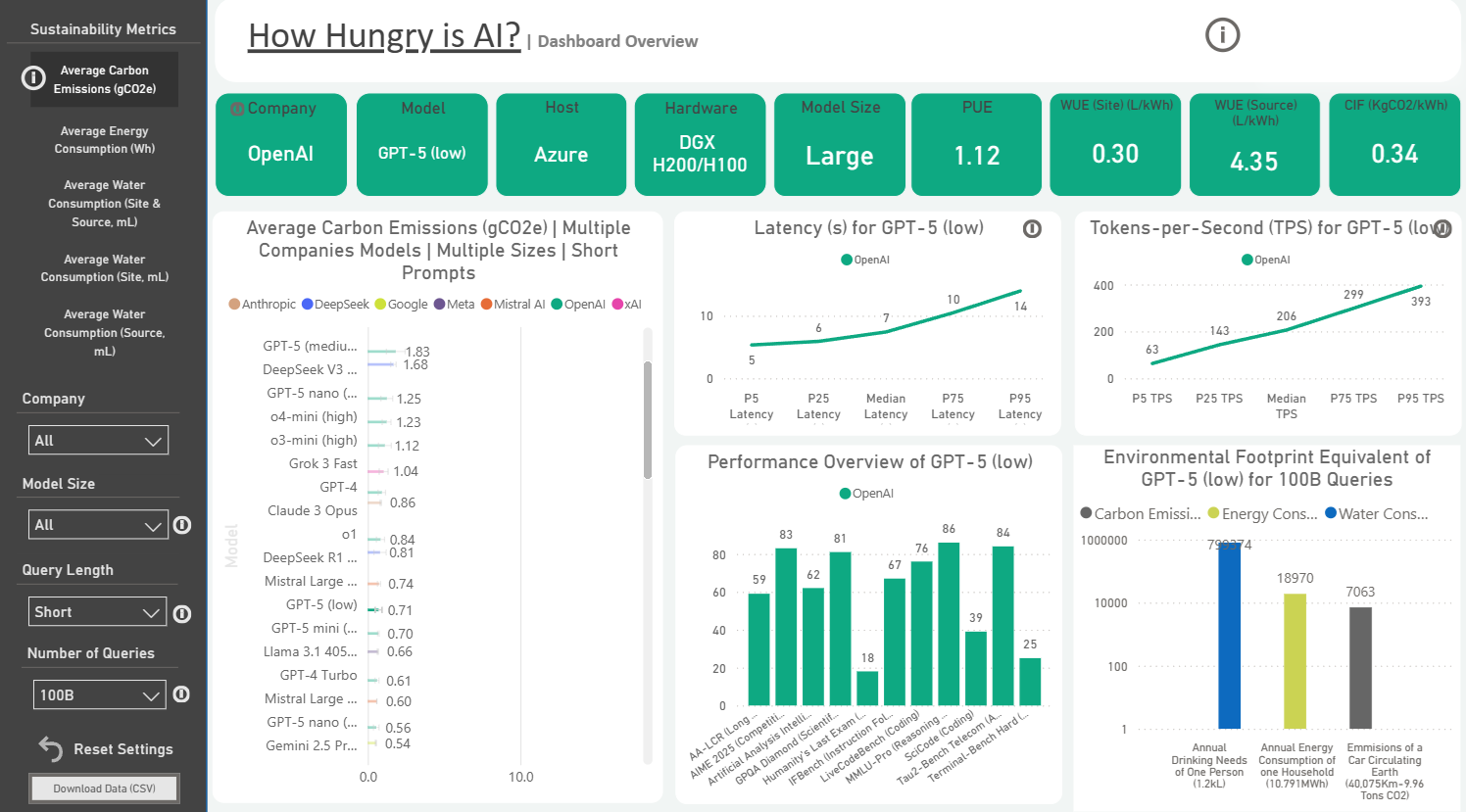}
        \caption{Overview of the main dashboard displaying the energy consumption per model, latency, TPS, benchmark scores, and equivalent environmental impacts for an example model (GPT-5 minimal).}
        \label{fig:dashboard_main}
    \end{subfigure}
    \hfill
    \begin{subfigure}[t]{0.48\linewidth}
        \centering
        \includegraphics[width=\linewidth]{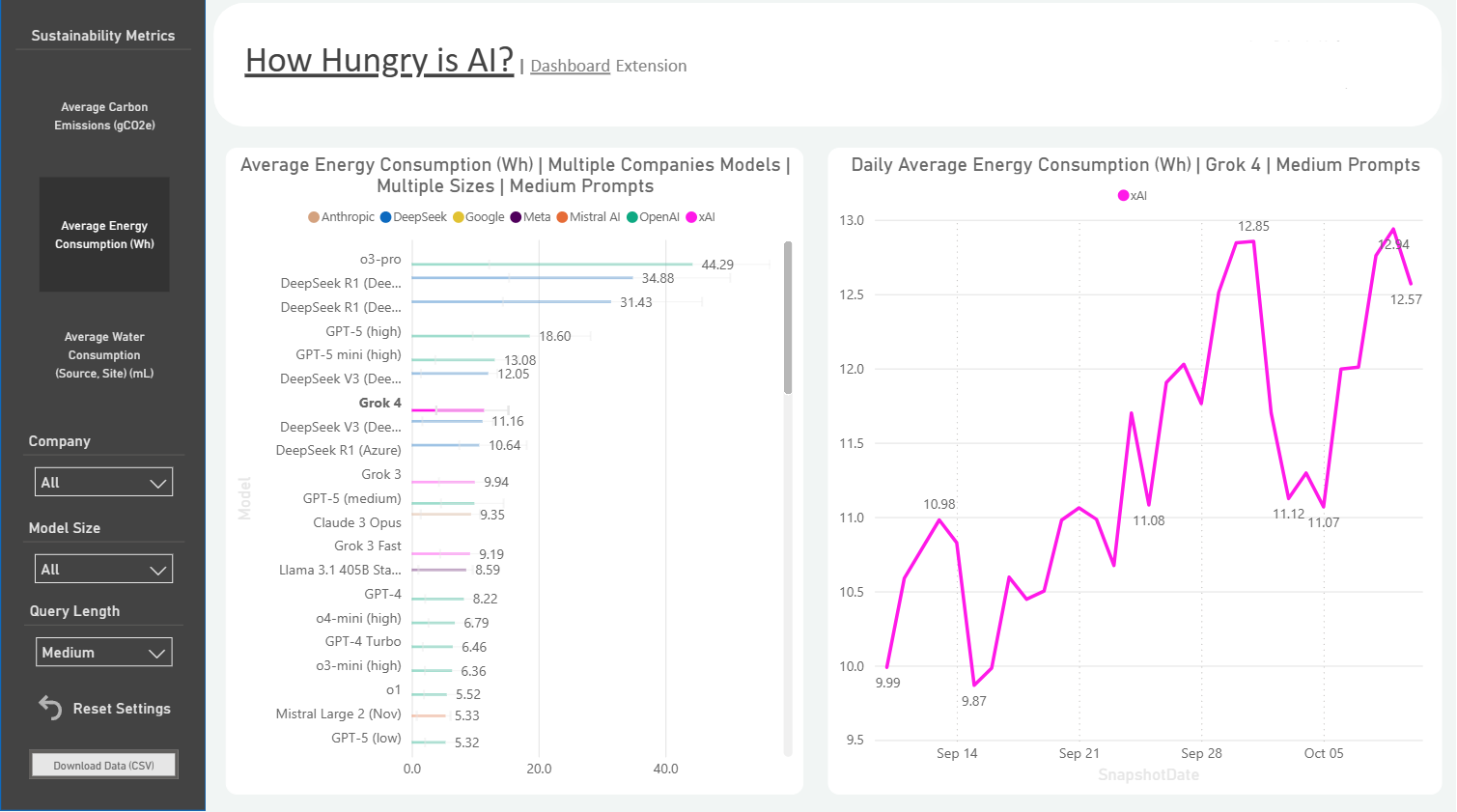}
        \caption{Overview of the timeseries dashboard displaying average energy consumption per model, and the daily fluctuations of the selected model (Grok 4).}
        \label{fig:dashboard_timeseries}
    \end{subfigure}
    \caption{Visual overview of the AI sustainability dashboard.}
    \vspace{-1em}

    \label{fig:dashboard_combined}

\end{figure}

To democratize access to these novel assessments, we built and deployed an automated Power BI dashboard that runs our entire framework in real time, a first-of-its-kind tool for continuously tracking AI inference sustainability. The data are scraped daily from the Artificial Analysis website, cleaned automatically, and then visualized on Power BI as seen in Figures~\ref{fig:dashboard_main} and~\ref{fig:dashboard_timeseries}. The main dashboard displays the average and standard deviation of energy use, water consumption (site, source, and combined), and carbon emissions for the three query sizes. It also visualizes latency and TPS fluctuations, benchmark results, and the total environmental impact when scaling up to $1$, $50$, or $100$ billion queries, compared with real-world equivalents such as household electricity use, annual drinking needs, and transportation emissions. Users can filter by company, model size, query size, or sustainability metric, and download the full dataset. Additionally, the dashboard tracks day-to-day changes in each model’s footprint, visualizing time-series trends and the average in energy, water, and carbon metrics across data centers and hardware setups. It includes an extended list of models beyond those analyzed in this study and automatically incorporates new ones as they are released, allowing continuous monitoring of inference-phase sustainability and cross-model comparisons over time.

\section{Experimental Evaluation} \label{sec:results}

We benchmark the environmental footprint of 30 LLMs across three modalities: Energy consumption, water usage, and carbon emissions, based on equations \ref{eq:eq3_4}, \ref{eq:water_cons}, and \ref{eq:carbon_emiss}, respectively. For the long-form query evaluation, GPT-4 and LLaMA-3 (8B and 70B) are excluded due to context window limitations.

\subsection{Energy Consumption}

\begin{figure}[htbp]
    \centering
    \begin{minipage}{0.49\textwidth}
        \centering
        \includegraphics[width=\linewidth]{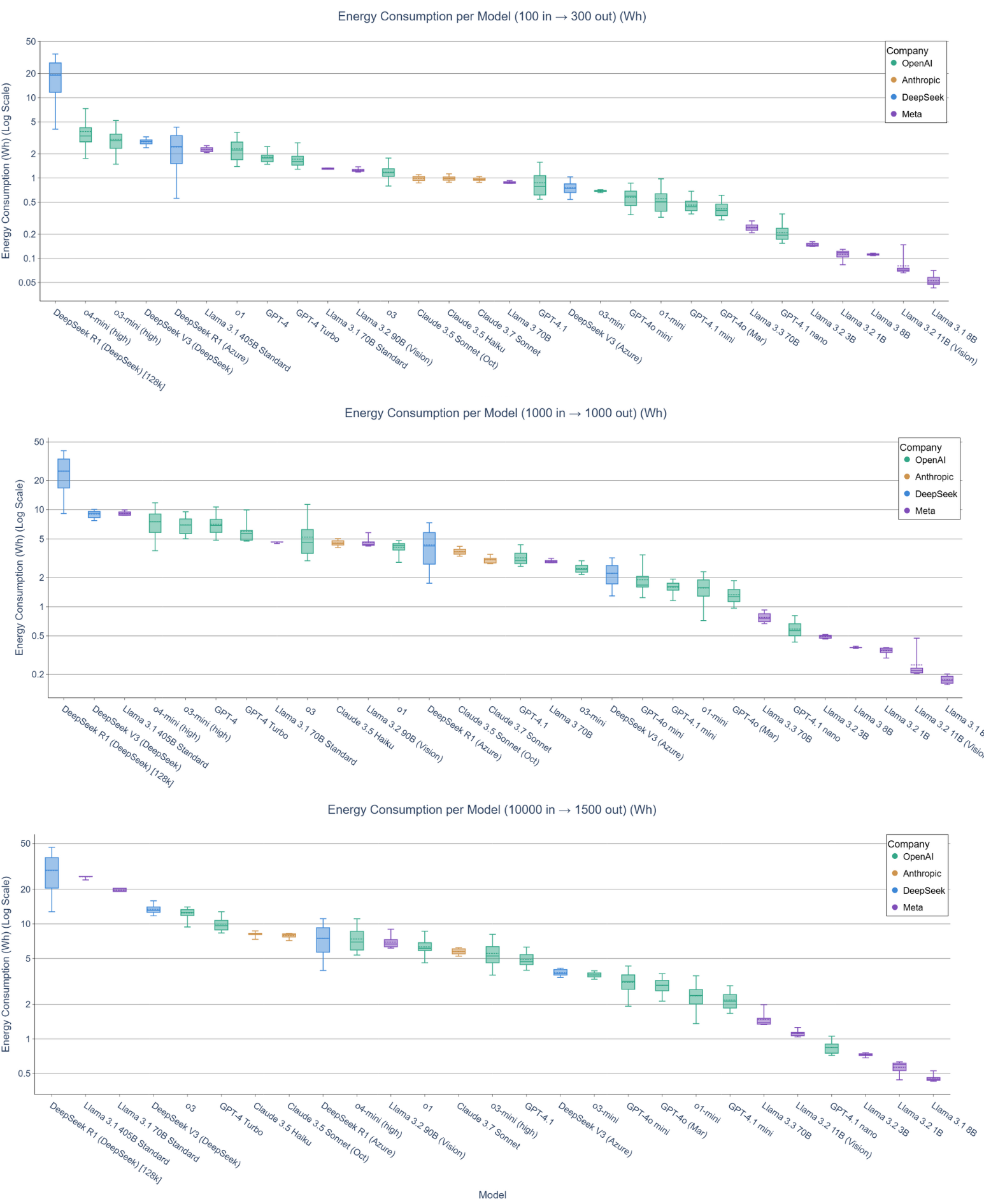}
        \caption{Energy consumption per model across three prompt sizes (Wh, log‐scale).}
        \label{fig:energy_graph}
    \end{minipage}
    \hfill
    \begin{minipage}{0.45\textwidth}
        \centering
        \captionof{table}{Energy consumption (mean ± std dev) per model across three prompt sizes (Wh).}
        \label{tab:energy_table}
        \resizebox{\textwidth}{!}{
       \begin{tabular}{cccc}
\hline
\textbf{Model}       & \textbf{\begin{tabular}[c]{@{}c@{}}Energy Consumption \\ (100 input-300 output)\\ (Wh)\end{tabular}} & \textbf{\begin{tabular}[c]{@{}c@{}}Energy Consumption \\ (1k input-1k output) \\ (Wh)\end{tabular}} & \textbf{\begin{tabular}[c]{@{}c@{}}Energy Consumption \\ (10k input-1.5k output) \\ (Wh)\end{tabular}} \\ \hline
GPT-4.1              & 0.871 ± 0.302   & 3.161 ± 0515   & 4.833 ± 0.650 \\
GPT-4.1 mini         & 0.450 ± 0.081   & 1.545 ± 0.211   & 2.122 ± 0.348 \\
GPT-4.1 nano         & 0.207 ± 0.047   & 0.575 ± 0.108   & 0.827 ± 0.094 \\
o4-mini (high)       & 3.649 ± 1.468   & 7.380 ± 2.177   & 7.237 ± 1.674 \\
o3                   & 1.177 ± 0.224   & 5.153 ± 2.107 & 12.222 ± 1.082 \\
o3-mini (high)       & 3.012 ± 0.991   & 6.865 ± 1.33   & 5.389 ± 1.183 \\
o3-mini              & 0.674 ± 0.015   & 2.423 ± 0.237   & 3.525 ± 0.168 \\
o1                   & 2.268 ± 0.654   & 4.047 ± 0.497  & 6.181 ± 0.877 \\
o1-mini              & 0.535 ± 0.182   & 1.547 ± 0.405   & 2.317 ± 0.530 \\
GPT-4o (Mar '25)     & 0.423 ± 0.085   & 1.215 ± 0.241   & 2.875 ± 0.421 \\
GPT-4o mini          & 0.577 ± 0.139   & 1.897 ± 0.570   & 3.098 ± 0.639 \\
GPT-4 Turbo          & 1.699 ± 0.355   & 5.940 ± 1.441   & 9.877 ± 1.304 \\
GPT-4                & 1.797 ± 0.259   & 6.925 ± 1.553   & --- \\
DeepSeek-R1 (DS) \textsuperscript{*}     & 19.251 ± 9.449  & 24.596 ± 9.4  & 29.078 ± 9.725 \\ 
DeepSeek-V3 (DS) \textsuperscript{*}     & 2.777 ± 0.223   & 8.864 ± 0.724   & 13.162 ± 1.126 \\
DeepSeek-R1 (AZ) \textsuperscript{†}    & 2.353 ± 1.129   & 4.331 ± 1.695  & 7.410 ± 2.159 \\
DeepSeek-V3 (AZ) \textsuperscript{†}     & 0.742 ± 0.125   & 2.165 ± 0.578   & 3.696 ± 0.221 \\
Claude-3.7 Sonnet    & 0.950 ± 0.040   & 2.989 ± 0.201   & 5.671 ± 0.302 \\
Claude-3.5 Sonnet    & 0.973 ± 0.066   & 3.638 ± 0.256   & 7.772 ± 0.345 \\
Claude-3.5 Haiku    & 0.975 ± 0.063   & 4.464 ± 0.283   & 8.010 ± 0.338 \\

LLaMA-3-8B           & 0.108 ± 0.002   & 0.370 ± 0.005   & --- \\
LLaMA-3-70B          & 0.861 ± 0.022   & 2.871 ± 0.094   & --- \\
LLaMA-3.1-8B         & 0.052 ± 0.008   & 0.172 ± 0.015   & 0.443 ± 0.028 \\
LLaMA-3.1-70B        & 1.271 ± 0.020   & 4.525 ± 0.053   & 19.183 ± 0.560 \\
LLaMA-3.1-405B       & 2.226 ± 0.142   & 9.042 ± 0.385   & 25.202 ± 0.526 \\
LLaMA-3.2 1B         & 0.109 ± 0.013   & 0.342 ± 0.025   & 0.552 ± 0.059 \\
LLaMA-3.2 3B         & 0.143 ± 0.006   & 0.479 ± 0.017   & 0.707 ± 0.020 \\
LLaMA-3.2-vision 11B & 0.078 ± 0.021   & 0.242 ± 0.071   & 1.087 ± 0.060 \\
LLaMA-3.2-vision 90B & 1.235 ± 0.054   & 4.534 ± 0.448   & 6.852 ± 0.780 \\
LLaMA-3.3 70B        & 0.237 ± 0.023   & 0.760 ± 0.079   & 1.447 ± 0.188 \\ \hline
\end{tabular}}

\begin{flushleft}
\tiny{\textsuperscript{*} DeepSeek Host} \newline
\tiny{\textsuperscript{†} Microsoft Azure Host}

\end{flushleft}
    \end{minipage}
\end{figure}

Figure~\ref{fig:energy_graph} and Table~\ref{tab:energy_table} highlight how energy consumption scales with prompt length and model architecture, revealing wide disparities across systems. LLaMA-3.1-8B is the most efficient, requiring only 0.443~Wh for long prompts (approximately 7,000 words of input and 1,000 words of output), followed by LLaMA-3.2~1B and LLaMA-3.2~3B at 0.552~Wh and 0.707~Wh, respectively. GPT-4.1~nano remains among the most efficient proprietary models at 0.827~Wh, but still consumes nearly twice the energy of LLaMA-3.1-8B. In contrast, DeepSeek-R1 (DS) consumes 29.075~Wh, around sixty five times more than the most efficient model, underscoring the large overhead of reasoning models. 

The LLaMA family shows clear scaling effects: energy use rises from 0.443~Wh at 8B parameters to 25.202~Wh at 405B, illustrating steep power demands at high parameter counts. Additionally, the DeepSeek models reveal striking infrastructure effects. DeepSeek-R1 and DeepSeek-V3 hosted on DeepSeek’s own servers consume 29.078~Wh and 13.162~Wh, while the same models on Azure use just 7.410~Wh and 3.696~Wh, over 70\% less energy. This gap highlights that hardware and data center efficiency, not model design alone, drives real-world energy use. For context, a single long query to DeepSeek-R1~(DS) consumes about as much electricity as running a 65-inch LED television ($\approx130\,\mathrm{W}$) for roughly 13 minutes. GPT-4o and GPT-4o mini also show that infrastructure can outweigh model size in determining energy efficiency. For instance GPT-4o consumes around 2.875~Wh while GPT-4o~mini's consumption is slightly higher at 3.098~Wh due to deployment on A100 hardware instead of H100s.

\subsection{Water and Carbon Emissions}
\begin{figure}[ht]
  \centering
  \begin{subfigure}[b]{0.49\linewidth}
    \centering
    \includegraphics[width=\linewidth]{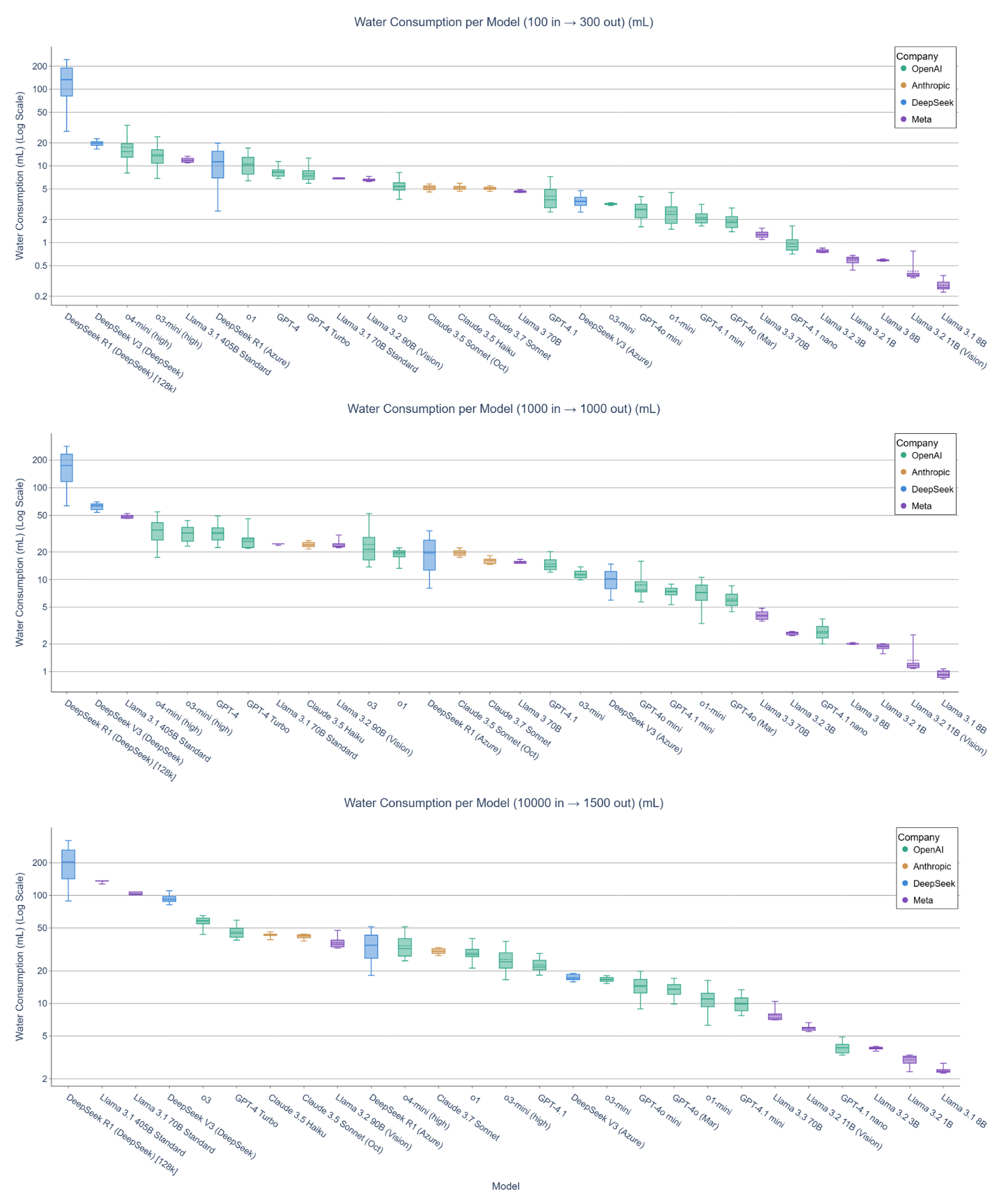}
    \caption{Water consumption per model across three prompt sizes (ml, log-scale).}
    \label{fig:water-total}
  \end{subfigure}\hfill%
  \begin{subfigure}[b]{0.49\linewidth}
    \centering
    \includegraphics[width=\linewidth]{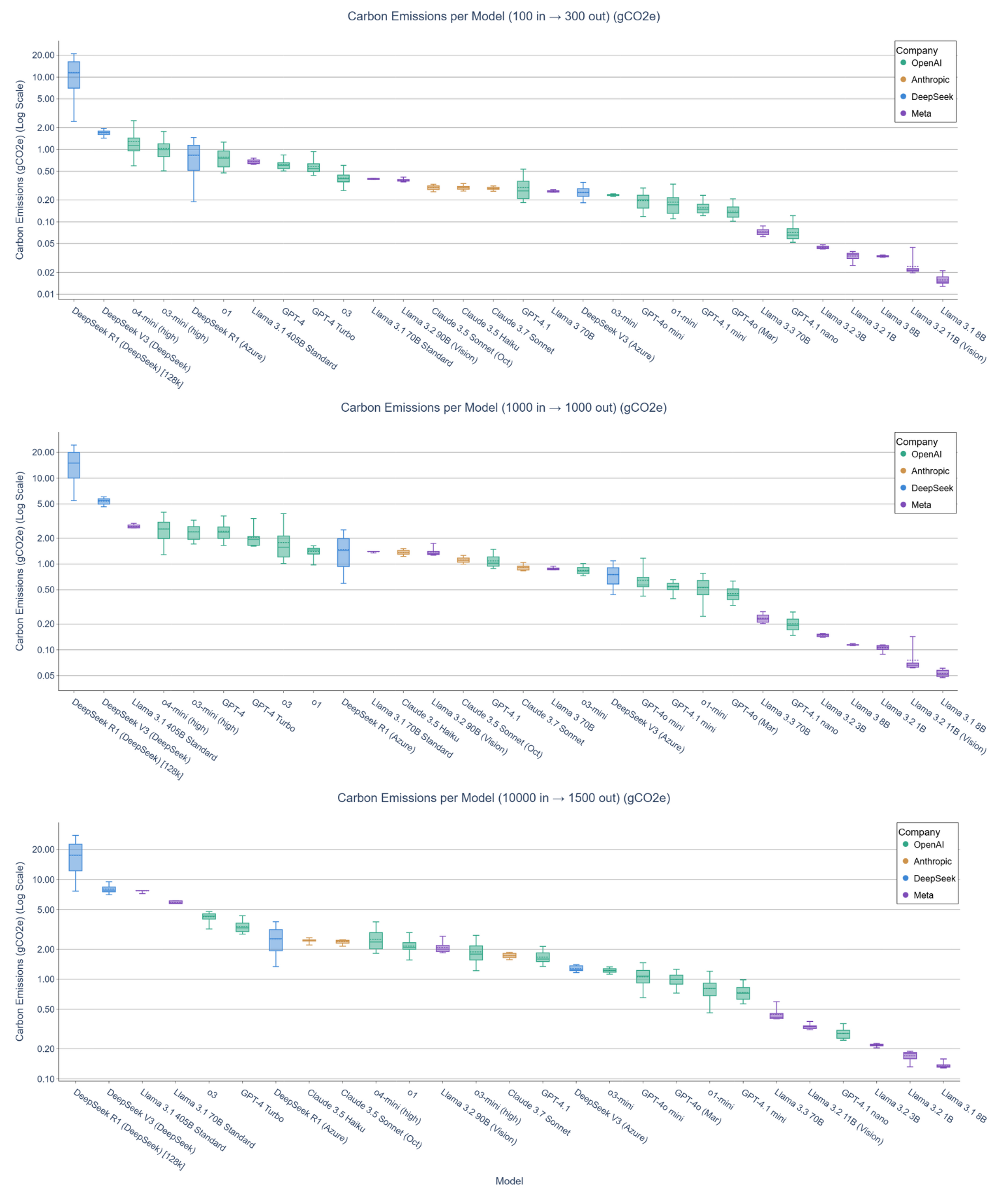}
    \caption{Carbon emissions per model across three prompt sizes (gCO$_2$e, log-scale)}
    \label{fig:carbon-total}
  \end{subfigure}
  \caption{Water consumption and carbon emissions per model.}
  \label{fig:carbon-vs-water}
\end{figure}

Figure~\ref{fig:carbon-vs-water} showcases the water consumption and carbon emissions of models across all prompt sizes. The most resource-efficient systems, such as LLaMA-3.2~1B, LLaMA-3.2~3B, LLaMA-3.1-8B, LLaMA-3-8B, and GPT-4.1~nano, emit less than 0.3~gCO$_2$e and consume under 4~mL of water even for long-form prompts, demonstrating exceptional sustainability across scales.

In contrast, large-scale and reasoning models such as o3, DeepSeek-R1 (DS), and DeepSeek-V3 (DS) exhibit substantially higher footprints. DeepSeek-R1 (DS) consumes over 200~mL of water and emits approximately 17~gCO$_2$e per long query, while the same model on Azure consumes only 34~mL and emits 2.5~gCO$_2$e, a reduction of nearly 85\%. These figures suggest that environmental impacts are shaped not only by model architecture but also by deployment strategies and regional infrastructure conditions. In particular, the elevated emissions and water usage observed in DeepSeek models likely reflect inefficiencies in their data centers, including higher PUE, suboptimal cooling technologies, and less efficient hardware.

While these per-query values may seem modest when isolated, their impact becomes considerable at scale. A single model, such as GPT-4o, serving hundreds of millions of daily requests, can emit as much carbon as thousands of transatlantic flights and consume water equivalent to the annual drinking needs of millions of people. We revisit this scaling analysis in greater detail in Section~\ref{section: Appendix E}.

\subsection{Validation Against Public Disclosures}
Public disclosures of inference-level energy and carbon data remain limited, but a few recent statements provide useful reference points for cross-validation. In June~2025, OpenAI CEO Sam~Altman reported that the default ChatGPT model consumed approximately 0.34~Wh per query~\cite{altman2025gentle}. Knowing that GPT-4o was the default deployment at that time, this estimate likely corresponds to GPT-4o-level inference. Our framework estimates 0.42~Wh ($\pm$0.13~Wh) for a short GPT-4o prompt (0.37~Wh without datacenter overhead), within 19\% of Altman’s figure. Similarly, the results for Mistral~Large~2 align closely with Mistral’s published life-cycle assessment (LCA) report \cite{mistral2025environmental}, which cites approximately 1.14~gCO$_2$e per 400-token query. Our corresponding estimate for 300 tokens (0.82~gCO$_2$e, $\pm$0.10~gCO$_2$e) scales to roughly 1.09~gCO$_2$e when normalized to 400 tokens, showcasing alignment within one standard deviation. Together, these alignments between independent disclosures and our modeled results suggest that the framework reproduces realistic operational conditions for modern LLM inference.

\section{GPT-4o Environmental Impact Case Study} \label{section: Appendix E}
\subsection{Energy Cost of a Single GPT-4o User Session}

\begin{figure}
    \centering
    \includegraphics[width=1
    \linewidth]{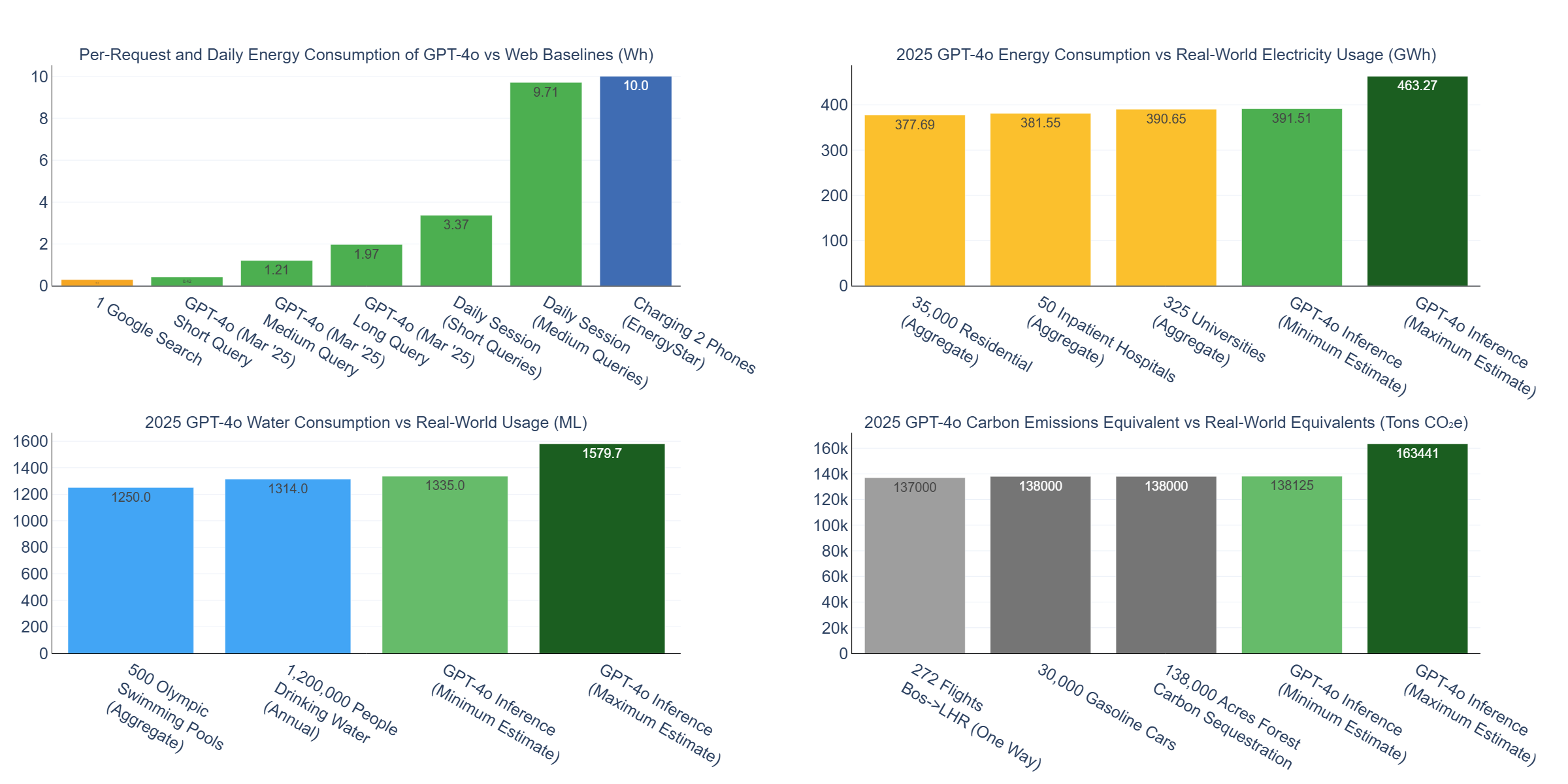}
    
\caption{(Top Left) Per-query and daily energy consumption of GPT-4o. (Top Right) Estimated total annual energy usage of GPT-4o in 2025. (Bottom Left) The estimated 2025 annual water consumption of GPT-4o. (Bottom Right) The estimated 2025 annual carbon emissions of GPT-4o.}
    \label{fig:gpt4ostudycase}
\end{figure}
Based on Reuters \cite{reuters2025openai}, the average ChatGPT user sends approximately eight queries per day as of April 2025. Based on this, we quantify the per-user energy impact of GPT-4o interactions against familiar digital activities as presented in Figure \ref{fig:gpt4ostudycase}. A single short GPT-4o query consumes 0.42 Wh (\(\pm 0.13\) Wh), exceeding the footprint of a Google search (0.30 Wh) by approximately 40\%. Scaling to a typical daily usage pattern, the cumulative energy reaches 3.73 Wh (\(\pm 0.358\) Wh). For medium-length queries, this increases to 9.71 Wh (\(\pm 1.106\) Wh). These results highlight that even limited daily engagement with GPT-4o can impose an energy cost comparable to charging two smartphones to full capacity (approximately 10 Wh), illustrating the tangible environmental footprint of conversational AI. While the individual per-query costs appear modest, their aggregation across millions of users introduces a rapidly compounding, largely invisible load on the environment.

\subsection{Estimated 2025 Annual Energy Consumption of GPT-4o Inference}

To estimate the annual energy demand of GPT-4o in 2025, we consider a baseline of 1 billion queries per day across all ChatGPT deployments, a figure reported by OpenAI as of December 2024 \cite{roth2024chatgpt}. Given GPT-4o's status as the default model, we conservatively attribute 700 million daily queries to GPT-4o. To simulate real-world usage dynamics, we apply a monthly prompt growth rate of 20\% from January to May 2025, reflecting the documented increase in ChatGPT’s weekly active user base from 300 million to 800 million between December 2024 and April 2025 \cite{singh2025chatgptstatistics}. This is followed by a decaying growth pattern from June to December, yielding a total of approximately 772 billion GPT-4o queries in 2025, which is around 15\% of the annual number of Google searches in 2024 \cite{cardillo2025google}. Within these queries, we conservatively assume an 80\%/20\% split between short and medium-length prompts based on typical usage patterns. Scaling the per-query energy estimates accordingly, we find that GPT-4o inference would require approximately 391,509 MWh annually at minimum and 463,269 MWh at maximum, as seen in Figure \ref{fig:gpt4ostudycase}. These values exceed the total electricity consumption of 35,000 U.S. residential households (377,685 MWh), 50 inpatient hospitals (381,550 MWh), and even 325 universities (390,650 MWh) annually. 

\subsection{Estimated 2025 Annual Water Footprint of GPT-4o Inference}
As showcased in Figure \ref{fig:gpt4ostudycase}, we translate estimated cooling and infrastructure-related water usage into real-world benchmarks. Based on scaled inference volumes, GPT-4o’s annual water consumption is projected to be between 1,334,991 kiloliters (kL) and 1,579,680 kL. These quantities are roughly equivalent to filling over 500 Olympic-sized pools or to supporting the annual drinking needs of 1.2 million people. Importantly, this consumption refers to evaporated freshwater permanently removed from local ecosystems rather than recycled. GPT-4o alone is responsible for evaporating an amount of freshwater equivalent to the annual drinking needs of almost 1.2 million people.

\subsection{Estimated 2025 Annual Carbon Footprint of GPT-4o Inference}

We further examine GPT-4o’s environmental footprint through estimated carbon emissions from electricity usage, as seen in Figure \ref{fig:gpt4ostudycase}. Our projections indicate annual emissions of approximately 138,125 tons of CO\(_2\)e at minimum and 163,441 tons at maximum. These figures are comparable to the annual emissions of 30,000 gasoline-powered cars or the cumulative emissions from approximately 272 transatlantic flights between Boston and London. In sequestration terms, offsetting GPT-4o’s annual emissions would require over 138,000 acres of average U.S. forest, an area roughly equivalent to the size of Chicago. These results showcase that the aggregation of hundreds of millions of requests per day can already impose a substantial environmental burden. This burden is only expected to grow as AI usage continues to scale.

\section{GPT-5 Adaptive Model Routing Case Study} \label{section: gpt-5}

The launch of GPT-5 \cite{openai2025introducinggpt5} introduced adaptive model routing, a mechanism that allows the system to automatically determine whether to use a fast variant or a more computationally intensive “Thinking” model for complex reasoning tasks. This unification eliminates the need for manual model selection where the model dynamically scales its reasoning effort based on prompt complexity.

\begin{figure}[H]
    \centering
    \includegraphics[width=\textwidth]{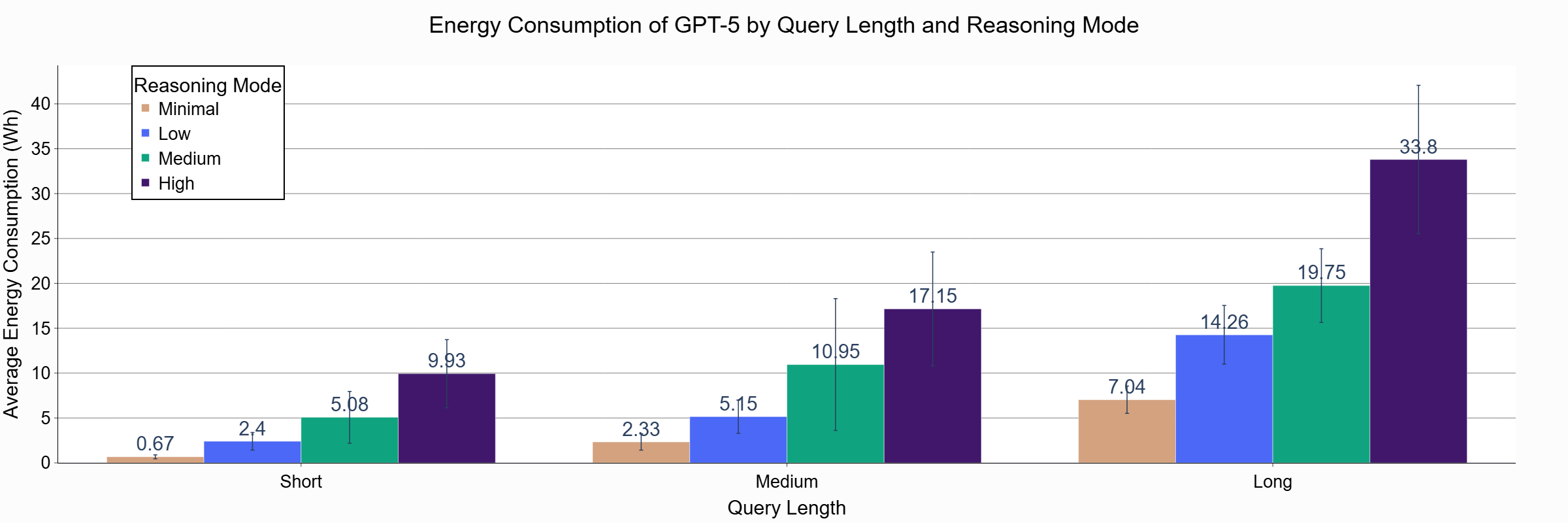}
    \caption{Energy consumption of GPT-5 across query lengths and reasoning modes}
    \label{fig:gpt5_reasoning}
\end{figure}

However, this adaptability introduces substantial variability in energy consumption across reasoning modes, as shown in Figure \ref{fig:gpt5_reasoning}. For medium-length queries, the average energy consumption ranges from 2.33Wh for minimal reasoning to 17.15Wh for high reasoning, representing a more than seven-fold increase. Despite this variance, GPT-5 remains relatively efficient at lower reasoning levels. For instance,  a short, minimal reasoning query consumes only 0.67 Wh, a value comparable to GPT-4o’s 0.42 Wh per short prompt. Conversely, a long, high-reasoning query reaches an average of 33.8 Wh, comparable to the upper bounds observed among the most energy-intensive models analyzed in this study.

These results suggest that while adaptive routing optimizes computational resources by tailoring inference depth to task complexity, it also amplifies the environmental footprint of cognitively demanding prompts. This finding underscores the growing importance of prompt-level efficiency analysis for next-generation LLMs that blend lightweight and high-reasoning architectures within a unified system.

\section{Discussion and Policy Implications} \label{sec:discussion}

\subsection{The Critical Role of Infrastructure in AI Sustainability}

Our findings indicate that infrastructure is a crucial determinant of AI inference sustainability. While model design enhances theoretical efficiency, real-world outcomes can substantially diverge based on deployment conditions and factors such as renewable energy usage and hardware efficiency. For instance, GPT-4o mini, despite its smaller architecture, consumes approximately 20\% more energy than GPT-4o on long queries due to reliance on older A100 GPU nodes. Similarly, DeepSeek models highlight the profound impact of infrastructure: DeepSeek-R1 and DeepSeek-V3 deployed on DeepSeek’s own servers exhibit water consumption and carbon emissions nearly six times higher than their Azure-hosted counterparts. The Azure deployments benefit from better hardware, more efficient cooling systems, lower carbon intensity, and tighter PUE control, demonstrating that sustainability gains can stem as much from datacenter design as from model optimization. These observations underscore that true AI sustainability will hinge on coordinated progress in hardware efficiency, renewable energy sources, and infrastructure-aware deployment strategies.

\subsection{Rebound Effects and the Jevons Paradox}

Although large language models consume significantly less energy, water, and carbon per task than human labor \cite{ren2024reconciling}, these efficiency gains do not inherently reduce overall environmental impact. As per-task efficiency improves, total AI usage expands far more rapidly, amplifying net resource consumption, a phenomenon aligned with the Jevons Paradox \cite{polimeni2008jevons}, where increased efficiency drives systemic demand. The acceleration and affordability of AI remove traditional human and resource constraints, enabling unprecedented levels of usage. Consequently, the cumulative environmental burden threatens to overwhelm the sustainability baselines that AI efficiency improvements initially sought to mitigate. As such, sustainable AI deployment must focus on systemic frameworks that assess how well models balance capability with environmental cost. In response, we propose DEA as a principled method for benchmarking model-level eco-efficiency.

\subsection{Policy Implications}

As AI systems scale globally, ensuring environmental sustainability requires both model-level optimizations and systemic regulation of infrastructure. Government agencies should encourage thresholds on the permissible environmental footprint per inference regarding energy, water, and carbon emissions that AI models must not exceed. These thresholds can be met through architectural innovations, such as sparsity and quantization, or through infrastructure-level optimizations like more efficient hardware, cleaner energy sourcing, and improved cooling systems. Our methodology offers a standardized, scalable framework to quantify these efforts. Incorporating technologies like dielectric liquid cooling offers a promising path to reduce or eliminate water use in data centers drastically \cite{10612781}. Transparency must also be elevated through system-level reporting of per-inference energy, water, and carbon metrics. Additionally, deployment strategies, such as batching, should be integrated into sustainability planning, as larger batch sizes can reduce per-query energy use by improving hardware utilization with only minimal impact on latency.

\section{Conclusion, Limitations, and Future Work} \label{sec:conclusion}

This paper introduces the first large-scale, infrastructure-aware framework for benchmarking the environmental footprint of LLM inference, integrating API performance, environmental multipliers, and statistical inference to assess energy, water, and carbon costs under real-world conditions. By applying cross-efficiency DEA, we contextualize environmental impact in terms of functional performance, revealing that eco-efficiency hinges not only on model design but also on infrastructure. Our GPT-4o case study emphasizes the Jevons Paradox: As AI becomes cheaper and faster, total usage expands, intensifying environmental strain despite gains in per-query efficiency. Additionally, our GPT-5 case study sheds lights on the importance of prompt-level efficiency and adaptive routing. Without structural shifts in how LLMs are designed, deployed, and used, these invisible costs will continue to rise, threatening to offset the societal benefits that made these systems valuable in the first place. This work establishes a standardized, scalable framework for benchmarking the environmental footprint of LLM inference in real-world data center deployments, providing a basis for transparent, infrastructure-aware sustainability assessment and future regulation.

Our work inherits certain limitations that we acknowledge: we avoid overstating model-specific footprints by conservatively including only the energy drawn by actively assigned GPUs. This is due to the lack of means to determine whether unused GPUs' capacity is reassigned, load-balanced, or left inactive. Isolating non-GPU power consumption was also difficult. We applied a fixed utilization estimate from prior studies, acknowledging that their variation across inference workloads is typically significantly lower than that of GPUs. Moreover, for proprietary models without disclosed size, we classified their scale based on observed API performance. Future work should address these limitations as more detailed telemetry and facility-level reporting become available. Additionally, future studies should also extend beyond text generation to evaluate image, video, and audio generation, which are likely to impose greater environmental costs due to higher computational intensity.


\newpage
\section*{Appendices}
\appendix

\section{Batch Size Sensitivity Analysis (GPT-4o)} \label{sec: Appendix B}

In our main analysis, we adopt a batch size of 8 for all per-prompt energy estimations. This choice reflects a middle ground in real-world deployments, where AI providers typically batch requests in the range of 4 to 16 to balance latency constraints with energy efficiency. However, the specific batch size used during inference can significantly influence energy consumption due to changes in GPU and system utilization.

To assess this effect, we present a sensitivity analysis using GPT-4o as a representative model. The only parameter varied is batch size, allowing us to examine how plausible batching configurations can significantly shift energy outcomes. This variation underscores the rationale behind our use of batch size 8 as a representative midpoint in real-world deployments.

\begin{figure}[h]
    \centering
    \includegraphics[width=1\linewidth]{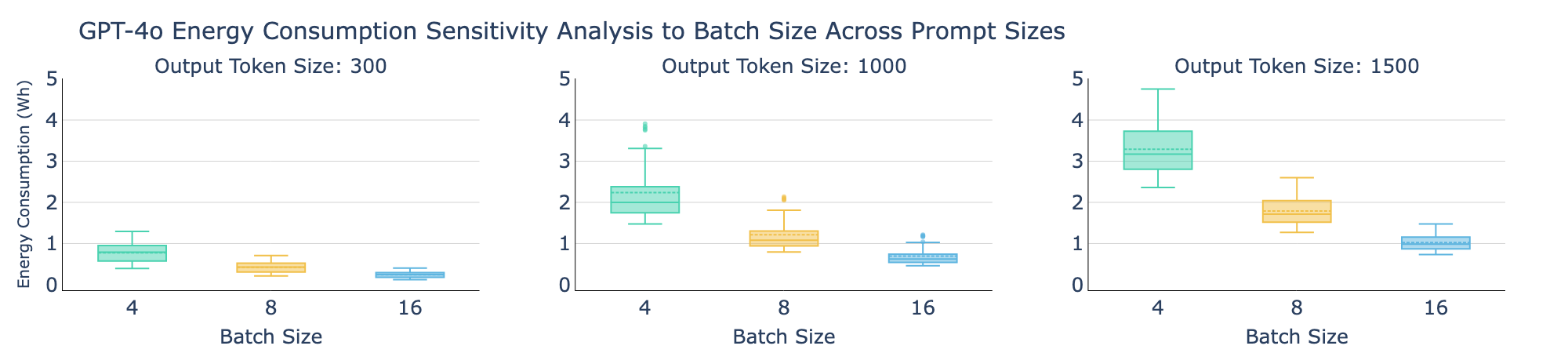}
    \caption{GPT-4o per-prompt energy consumption (Wh) across batch sizes and prompt lengths.}
    \label{fig:batch_sensitivity_plot}
\end{figure}
Table \ref{tab:utilization_assumptions} summarizes the utilization rates applied to each batch size, following the same method used in our methodology section \ref{sec:methodology}, which drives the corresponding per-prompt energy estimates shown in Figure \ref{fig:batch_sensitivity_plot}.

\begin{table}[]
    \centering

\caption{Estimated node-level GPU and non-GPU utilization by batch size for GPT-4o.}
\vspace{0.5em}
\begin{tabular}{cccc}
\hline
Batch Size & $D_{\text{GPU}}$ & $U_{\text{GPU total}}$ & $U_{\text{non-GPU total}}$ \\ \hline
4          & 40-55\%          & 10-13.5\%              & 12.5\%                     \\
8          & 45-60\%          & 5.5-7.5\%              & 6.25\%                     \\
16         & 55-70\%          & 3.5-4.5\%              & 3.125\%                    \\ \hline
\end{tabular}
\label{tab:utilization_assumptions}
\end{table}

The results show substantial efficiency gains with higher batching: moving from batch size 4 to 8 reduces energy per prompt by approximately 45\%, while increasing from 8 to 16 yields a further 43\% reduction. If we had used a batch size of 4 throughout our study, energy estimates would have been significantly higher, overstating the environmental footprint of LLM inference. Conversely, using a batch size of 16 would have resulted in notably lower energy values, possibly underestimating the footprint in more latency-constrained or low-traffic scenarios.

These differences highlight the critical role that batching decisions play in shaping the environmental footprint of large-scale LLM deployments. As AI models utilize dynamic batching to address traffic and latency issues, adjusting the batch size can significantly impact the environmental footprint of each prompt. Large-scale providers like OpenAI have a significant advantage in this regard, as their high traffic volume allows them to rely on higher batch sizes without sacrificing latency to the same extent as smaller or less active deployments.

\section{Scope 3 Considerations} \label{sec: Appendix A}

While this study focuses on operational emissions and resource consumption during inference (Scopes 1 and 2), it is important to briefly discuss the Scope 3 impacts associated with the manufacturing, transportation, and end-of-life disposal of the hardware used to power LLMs.

Scope 3 emissions are typically the most significant contributor to the lifecycle footprint of data center infrastructure, encompassing embodied carbon from GPU fabrication, water usage in semiconductor manufacturing, emissions from global logistics, and hardware retirement. For instance, Microsoft's Scope 3 CO$_2$e emissions in 2023 accounted for 66\% of the total emissions \cite{microsoft2024environmental}. Yet, these values are highly variable across vendors, manufacturing locations, and fabrication nodes, and they lack deployment-specific attribution when applied to real-time inference tasks.

Moreover, given that many large-scale models are continually updated and deployed across evolving infrastructures, ascribing a fixed fraction of embodied emissions or water per query is both methodologically fragile and likely to result in overestimation. Applying complete hardware manufacturing footprints to ongoing inference, without amortizing them over the expected hardware lifespan or query volume, risks artificially inflating per-query environmental costs.

In light of this, we excluded Scope 3 from our prompt-level framework, as its inclusion would introduce non-trivial uncertainty and potentially distort comparative eco-efficiency across models. Nevertheless, the long-term sustainability of AI infrastructure will depend on extending lifecycle accountability beyond the inference phase; future work is encouraged to adopt comprehensive lifecycle analyses (LCA) that integrate Scope 3 considerations once transparent and standardized data become available.

\begin{figure}
    \centering
    \includegraphics[width=\linewidth, trim=0 32 0 0, clip]{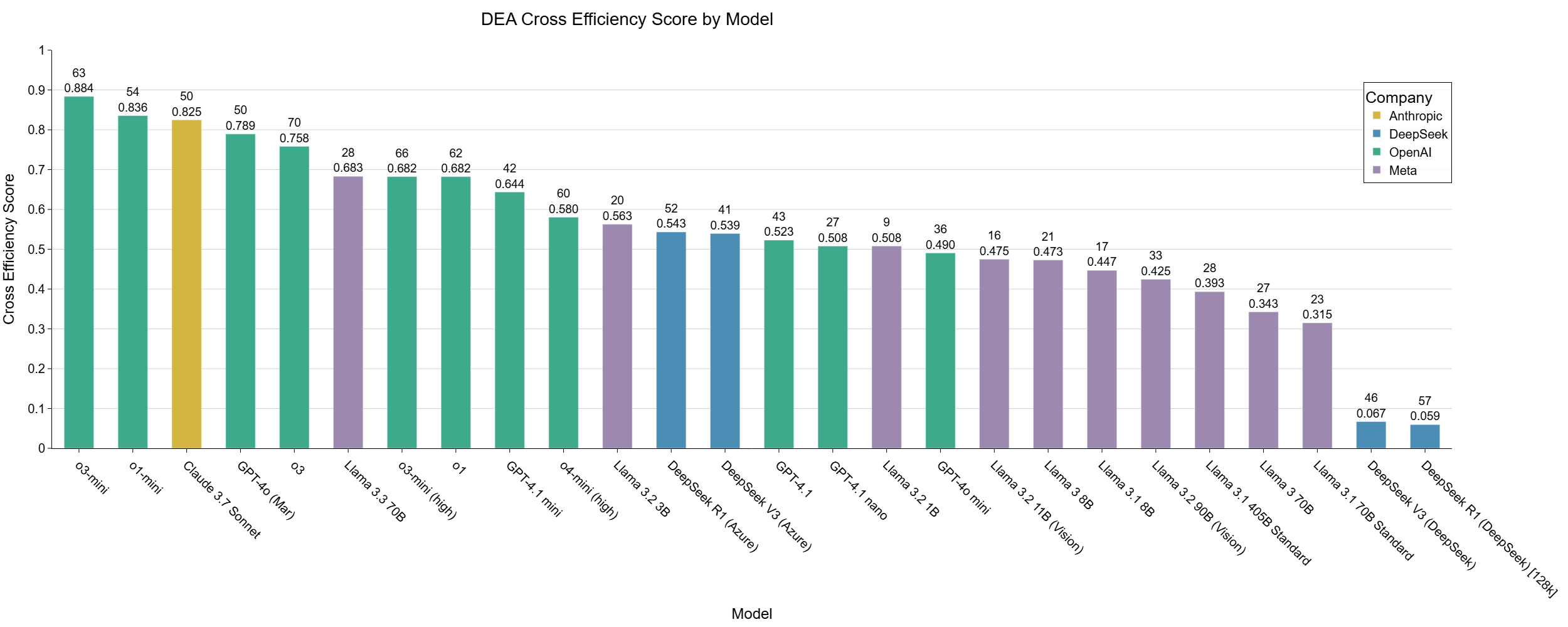}
    \caption{Cross efficiency DEA scores. Bar labels show the AI Index (top) and cross-efficiency score (bottom).}
    \label{fig:dea}
\end{figure}
\section{Cross-effficiency DEA Results} \label{sec: Appendix C}
Before presenting the eco-efficiency results, it is worth noting that Claude 3.5 Sonnet, Claude 3.5 Haiku, GPT-4, and GPT-4 Turbo were excluded due to the lack of benchmark results on certain tests. Since cross-efficiency requires complete inputs and outputs, these models could not be fairly evaluated.

As shown in Figure~\ref{fig:dea}, OpenAI’s reasoning models dominate the eco-efficiency frontier. o3-mini achieved the highest cross-efficiency score (0.884), closely followed by o1-mini (0.836) and Anthropic’s Claude 3.7 Sonnet (0.825), which combines strong reasoning ability with a relatively modest environmental footprint. GPT-4o (Mar) (0.789) and o3 (0.758) also performed well. These results suggest that downsizing reasoning models can yield meaningful sustainability gains without compromising performance.

At the opposite end, DeepSeek-R1 (0.067) and DeepSeek-V3 (0.059) recorded the lowest efficiency scores. Despite their advanced reasoning capabilities, their high energy, water, and carbon costs indicate significant infrastructural inefficiencies. Their Azure-hosted variants performed better, DeepSeek-R1 (0.539) and DeepSeek-V3 (0.523), yet remained below most OpenAI and Anthropic systems. Among OpenAI models, GPT-4.1 mini (0.580) and GPT-4.1 nano (0.508) balanced output quality and sustainability particularly well. LLaMA models clustered between 0.4 and 0.6, reflecting efficient power use but limited reasoning performance. 

In summary, eco-efficiency relies on both output quality and environmental cost. OpenAI’s smaller reasoning models and Claude 3.7 Sonnet strike that balance most effectively, while DeepSeek and LLaMA demonstrate the limitations of concentrating on capability or sustainability alone.

\end{document}